**ORIGINAL RESEARCH ARTICLE**

**TITLE:**  Examining Online Social Support for Countering QAnon Conspiracies

**Michael Robert Haupt, PhD\*[1,2,3], Meng Zhen Larsen, BA, BS\*[3,5], Michelle Strayer, MA[4], Luning Yang[3], Tim K. Mackey, MAS, PhD[2,3,5]**

***\* Co-first authors***

[1] Department of Cognitive Science, University of California San Diego, 9500 Gilman Dr, La Jolla, CA, 92093, USA
[2] Global Health Program, Department of Anthropology, University of California, San Diego, CA USA
[3] Global Health Policy & Data Institute, San Diego, CA USA
[4] Columbia University, New York, NY USA
[5] S-3 Research, San Diego, CA USA

**Word Count:**  8,938

**Tables/Figures:**  6 Tables,  1 Figure

**Address for Correspondence:** Michael Robert Haupt, MA

Department of Cognitive Science
9500 Gilman Dr
La Jolla, CA, 92093, USA
**Email**: mhaupt@ucsd.edu

## Abstract

As radical messaging has proliferated on social networking sites, platforms like Reddit have been used to host support groups, including support communities for the families and friends of radicalized individuals. This study examines the subreddit r/QAnonCasualties, an online forum for users whose loved ones have been radicalized by QAnon. We collected 1,665 posts and 78,171 comments posted between 7/2021 and 7/2022 and content coded top posts for prominent themes. Sentiment analysis was also conducted on all posts. We find venting, advice and validation-seeking, and pressure to refuse the COVID-19 vaccine were prominent themes. 40% (n=167) of coded posts identified the user's Q relation(s) as their parent(s) and 16.3% (n=68) as their partner. Posts with higher proportions of words related to swearing, social referents, and physical needs were positively correlated with engagement. These findings show ways that communities around QAnon adherents leverage anonymous online spaces to seek and provide social support.

*Keywords*: Online Support Groups; QAnon; Conspiracies; Anti-Vax; Sentiment analysis; Reddit

## Introduction

With the growth of communication technologies over the past few decades, online platforms are increasingly used as mediums for personal expression (Bailey et al., 2020), creating and maintaining social connections across geographic distance (Scellato et al., 2010), and large-scale coordination efforts such as protests and astroturfing (Haupt, Jinich-Diamant, et al., 2021; Haupt, Xu, et al., 2021). Unfortunately, this freedom of expression and access to a wide variety of content has drawbacks: newsfeed algorithms have been shown to shape media consumption on social networking sites (Cohen, 2018), marginalize moderate media (Benkler et al., 2018), and increase the proliferation of extremist ideologies (Forberg, 2022; Lim, 2017).

In recent years, social media platforms have also influenced public discourse surrounding events like the 2016 US Presidential election (Carlson, 2020) and the COVID-19 pandemic, including hosting misinformation and conspiracy theories (Napoli, 2019; Zarocostas, 2020).The vast majority of social networking sites do not regulate or vet content, instead adhering to a set of loose community guidelines allowing users to generate and share content with little requisite oversight (McChesney and Pickard, 2011).  Social media platforms have consequently proliferated mis-, dis-, and mal-information ranging from organized international attacks  to COVID-19 conspiracy theories (see, for example: the US Department of State's GEC Special Report on Russian disinformation strategy: "Russia's Pillars of Disinformation and Propaganda," 2020; Haupt et al., 2023; Haupt, Li, et al., 2021).

Recent research offers several competing theories regarding the causes of conspiracism and conspiracy beliefs. Some research argues that at its root, conspiracy ideation is driven by the need for cognitive closure amidst uncertainty (Marchlewska et al., 2018). Other theories suggest that growing disaffection with modern political institutions drives some individuals to challenge authoritative news and media sources and turn toward counternarratives and conspiracies to understand the world and their role within it (Harambam and Aupers, 2015; Mahmud, 2022). Additional research has also focused on the impact of social motives, suggesting conspiracies allow individuals to feel unique and special by having exclusive information, thus bolstering in-group identities by denigrating outgroups (Douglas et al., 2019). Unfortunately, conspiracy groups and political extremists take advantage of this conspiracy psychology and the modern media landscape (Nichols, 2017), using online platforms to proliferate disinformation and conspiracy narratives.

Of course, political extremism and conspiracy theories long predate social networking sites and online media platforms. Before the advent of the internet, conspiracy theories and extremist messaging proliferated via newspapers, radio, television, books, pamphlets, and newsletters. The American media landscape saw several documented spikes in conspiracism over time, including around 1899 during the second industrial revolution and the late 1940s/early 1950s during the beginning of the Cold War (Uscinski and Parent, 2014). In a more recent example, the infamous 1978 novel *The Turner Diaries,* a thinly veiled treatise and instructional on white supremacy and race war, has been credited with radicalizing extremists like Timothy McVeigh, as well as inspiring acts of terrorism and the formation of white supremacist and paramilitary groups. Nevertheless, modern social media platforms represent new avenues for the dissemination of extremist content, conspiracy theories, and online radicalization. Many extremist and conspiracy groups, including QAnon, continue to use online spaces to attract membership, proliferate content, and organize.

**QAnon Definition and Brief History**

QAnon is an online conspiracy group that originated on the forum site 4chan in 2017. The group follows a decentralized set of conspiracy beliefs, principally alleging that the United States government has been infiltrated by a cabal of Satan-worshipping pedophile politicians (Roose, 2021). Followers of QAnon believe the movement is led by one or more high-ranking government officials with ultra top secret "Q-level" security clearances, who are associated with former U.S. President Donald Trump (Papasavva et al., 2022). QAnon entered mainstream alt-right conservative politics after the 2016 "Pizzagate" incident, when a series of leaked presidential campaign emails and spiraling online conspiracy theories led to a violent confrontation at a DC pizza restaurant.[1]

After Pizzagate, QAnon gained traction in alt-right conservative discourse with the emergence of "Q", an online personality who claimed to be a government official with ties to the Trump administration. Q created a post on 4chan titled "The Calm Before the Storm," alleging that former President Trump planned to initiate an attack against so-called "deep-state" cabal actors. This initial post, called a "Qdrop," was followed by thousands of similar posts in the ensuing years, drawing an audience and building an overarching conspiracy movement (Hodwitz et al., 2022).

The Qdrop conspiracy structure encourages interactivity. QAnon adherents form online communities dedicated to decoding Qdrops—believed to be breadcrumbs directing users to some ultimate truth about the world and existing political institutions (Frincu, 2021). Rather than enforcing a single rigid ideology, this structure allows individuals to interpret the Qdrop breadcrumbs and draw their own varied conclusions. QAnon also uses memes, far-right iconography, and pop-culture references to engage the Q community and proliferate content. This dynamic creates a shared social world on online forums and media platforms through shared memes, in-group conspiracy references, and tailored jokes (Daniels, 2018). The Q model thus combines breadcrumb Qdrop posts with shared subculture lexicons, promoting user collaboration as followers uncover the "truth."

---

[1]During the runup to the 2016 US Presidential election, emails from John Podesta, former presidential candidate Hilary Clinton's campaign chairman, were leaked online. Users on online forums like 4chan, including Russian trolls, speculated the Clinton campaign was secretly distributing child pornography. The conspiracy theory evolved in the ensuing weeks, claiming members of the US Democratic party ran a to include cannibalistic child sex trafficking ring in the basement of a DC pizza parlor (Aisch et al., 2016). Pizzagate culminated when a conspiracist with AR-15 firearms drove to the pizza parlor and held staff and patrons hostage in an attempt to uncover the alleged ring (Hsu, 2017).

**Online Community Support Groups**

In response to the rise of social networking sites, previous studies have explored the utility of online support groups. Often using forums, chat rooms, or social media platforms, online support groups are easily accessible 24/7, requiring only an internet-connected device (Vilhauer et al., 2010). Like in-person groups, online support groups draw individuals who share similar difficulties, on the premise that those with similar lived experiences may better understand each other and offer advice or emotional support. Previous research suggests support groups can improve participants' mental wellbeing (Setoyama et al., 2011), empowerment (van Uden-Kraan et al., 2009), and self-efficacy (Finn et al., 2009); and, like face-to-face groups, online support groups help provide catharsis, mutual support, empathy, and problem solving tools (Finn, 1999).

In comparison to physical spaces, the features of web-based environments offer different situational contexts that influence group dynamics. To explore how these web-based features may influence group dynamics, we adopt an affordance lens. An affordance, in this case, refers to a relational framework of technology use that accounts for dynamics between users, features of technology, and how the technology is used (Evans et al., 2017; Norman, 1988).[2] An example

of an affordance that exists across multiple social networking sites is *support giving*, which can

be expressed when users use upvote or like features on a post (Hong et al., 2017; Moreno and

Uhls, 2019; Syrdal and Briggs, 2018). Unlike in-person groups, online support groups on

platforms like Reddit offer anonymity by hiding user identity and personal information behind

avatars and usernames. Without the need to show their face or disclose their identity, users may

feel more comfortable sharing personal or sensitive information and expressing themselves

openly—a phenomenon called the *online disinhibition effect* (Pedersen, 1997; Suler, 2004).

As suggested in previous work, the online disinhibition effect may have a positive effect on the success of support forums as users, to some extent, must feel comfortable expressing themselves and interacting with other members (Barak et al., 2008; Clark-Gordon et al., 2019;

Joinson, 2001; Lapidot-Lefler and Barak, 2015; Tanis, 2007).  It is therefore likely that web-

based spaces that have affordances promoting a disinhibition effect may further increase participation and quality of support given in these groups. Existing research indicates additional affordances associated with the online disinhibition effect in communication platforms such as invisibility and delayed reactions (Barak et al., 2008).

Similar yet distinct from anonymity, invisibility refers to the aspect of online communication where users cannot physically see each other, creating less concern for how someone sounds or looks (Suler, 2004). For support groups that deal with socially stigmatized topics,  the anonymity and invisibility of online communication may allow participants to be more open in expressing themselves or candidly discussing their concerns (Davison et al., 2000). Additionally, the lack of physical social cues (e.g., someone shaking their head, looking bored, etc.) or status/authority cues (e.g., associated with a person's clothes) may also impact interactions. Online support groups allow individuals to bypass these cues, which may further promote the disinhibition effect (Suler, 2004).

---

[2] See the following for further examples of studies that apply an affordance lens when examining online behaviors such as drug selling tactics, cyberbullying, and social comparisons: (Chan et al., 2019; Fox and Moreland, 2015; Haupt et al., 2022)

Most online forums, including Reddit, also allow a delay in receiving responses and reactions from other users. Unlike in-person communication, where sharing a story may be immediately met with a nod or a sigh of disapproval, the lack of real-time feedback may lead individuals to disclose more information (Taylor and Macdonald, 2002). Additionally, delayed reaction times and the ability to edit posts allow users to choose exactly how much information to disclose, which may be of particular importance for individuals uncomfortable with in-person communication (Tsan and Day, 2007) or who prefer more thorough deliberation in their responses (Walther and Boyd, 2002).

**Present Study**

Given the potential of online communities to provide anonymous support to those in need, this paper examines Reddit discourse and user engagement on the group r/QAnonCasualties. QAnonCasualties is a forum for users struggling to cope with a friend or loved one's involvement in QAnon. QAnonCasualties offers a space for "support, resources, and a place to vent," where users can "heal themselves" and "learn how to steer [their loved ones] back to reality," according to the group's description as of August 2024 (QAnonCasualties, n.d.). The support group has over 273,000 members and has been reported on by popular press (Carrier, 2021; Jackson, 2021). While existing scholarship examines virtual spaces where Q-related conspiracies are disseminated (Bleakley, 2023; Engel et al., 2022; Papasavva et al., 2021, 2022; Sipka et al., 2022), there is currently no work that formally characterizes discourse and user engagement in support groups such as r/QAnonCasualties for those who have had friends, family and other acquaintances radicalized by QAnon.

To fill this gap, this research mined and content coded high engagement posts to identify prominent themes on r/QAnonCasualties between July 2021 and July 2022. We also conducted sentiment analysis and bivariate correlations to examine the relationships between emotional post content and engagement behaviors (upvoting and commenting). By identifying discussion themes among high-engagement posts, this paper suggests areas of need among the concerned families, friends, and loved ones of QAnon adherents. Findings from this study aim to inform efforts by mental health professionals, organizations, and similar forums that provide support and resources to the communities of concern around individuals engaged with QAnon.

## Methods

### Data Collection

A total of 1,665 posts and 78,171 user comments were collected from the QAnonCasualties subreddit from July 15th 2021 to July 15th 2022 using the then-available Reddit PushShift API (Version 0.1.2). The number of comments and upvotes for each post were combined into a single engagement score. Given that commenting is more effortful, requiring users to generate a written response (Haupt, Xu, et al., 2021), comments were weighted twice as high as upvotes. Posts in the 75th percentile of engagement were selected for inductive content coding to identify prevalent discussion themes.

**Inductive Content Coding**

Authors 1, 2, and 3 each coded the same initial 50 posts to develop a preliminary codebook of themes. The authors then met to converge theme categories and refine theme codes and sub-codes. Once the codebook was complete, the authors separately coded the remaining posts. Identified themes included: mention of conspiracies (general or COVID-related), pressures about not receiving COVID vaccinations, the political orientation of the Q relation, the religious affiliation of the Q relation, whether the Q relation got COVID, whether the user was still in contact with the Q relation (Disconnect), ways in which the user believed the Q relation was radicalized, and mentions of the Q relation becoming deradicalized. Posts were also classified by function: whether the user sought validation/corroboration, advice, or a space to vent about their experiences. Mentions of relationship types (e.g., parent, sibling, friend) of the Q relation to the user were coded as well.

**Sentiment Analysis**

Sentiment analysis was run on all 1,665 posts and 78,171 comments. Sentiment scores were calculated using Linguistic Inquiry and Word Count (LIWC), a text analysis program that reflects the percentage of words within a post that correspond to a given sentiment category. LIWC has been used in previous research investigating sentiments of conspiracy discourse on social media (Fong et al., 2021; Haupt et al., 2023; Rains et al., 2021) and characterizing post content within online support groups (Alpers et al., 2005; Han et al., 2008; Wang et al., 2015).

The sentiment scores were calculated to assess language related to: Analytic thinking (metric of logical, formal thinking), Clout (language of leadership, status), Authenticity (perceived honesty, genuineness), and Netspeak (internet slang), as well as cognitive processes related to information evaluation (i.e., Causation, Discrepancy, Tentative, Certitude),  Emotional Affect (i.e., Affect, Positive Tone, Negative Tone, Emotion, Positive emotion, Negative emotion, Anxiousness, Anger, Sadness), social and health-related topics (i.e., Prosocial behavior, Interpersonal conflict, Moralization, Health, Illness, Death, and Risk), and time-related sentiments (i.e., Time, Past focus, Present focus, Future focus). Sentiments that only referred to the text-related properties of a post and not its semantic content (i.e., word count, dictionary recognized words, words per sentence, etc.) were removed from analysis.

Pearson bivariate correlations were run to assess potential associations between LIWC sentiment variables and the number of upvotes and comments posts received. To examine the sentiments most commonly used by high-engagement users, the average number of comments was calculated per user account, identifying the top 10% of all commenters. Sentiment scores were averaged across comments from top commenter accounts and the resulting means were then ranked.

## Results

### Content Coding Themes

Among the posts that received the most engagement (n=418), as seen in **Table 1**, 61.5% (n=257) involved venting about experiences with QAnon family members or friends, about 19.6% (n=82) were advice-seeking, and 7.4% (n=31) looked to validate experiences. COVID-19 surfaced as a popular theme. More than half of the coded posts (60.5%, n=253) mentioned the Q relation discussing conspiracies related to the COVID-19 vaccine, and 37.6% (n=157) of post authors discussed being pressured to refuse the COVID-19 vaccine, and/or being pressured to use alternative (and scientifically unsupported) treatments like Ivermectin. Additionally, 3.8% (n=16) of posts showed users seeking reassurance about receiving the vaccine after discussing frightening anti-vax theories with their Q relation(s). About 0.7% (n=3) of posts included reassurances about getting vaccinated. Despite wanting to receive the vaccine, users described "living in paranoia and fear," expressing uncertainty about the safety of the vaccine and fear about the potential repercussions from their Q relation(s) if they chose to be vaccinated.

In posts where the political orientation of the Q relation was mentioned, 14.4% (n=60) identified their Q relation as conservative, compared to 1.9% (n=8) identified as liberal.[3] Religious affiliation was not mentioned often (only 7.9% of posts, n=33); however, when discussed, 31 out of 33 posts described the Q relation as Christian (7.4% of overall posts, n=31). When posts mentioned COVID infection status, 8.1% (n=34) stated their Q relation contracted COVID-19 or passed the virus to others. About 2.2% (n=9) stated their relation had passed away from COVID-19.

Of the top posts, 21.8% (n=91) discussed the way they believed their Q relation was radicalized. The top cited method of radicalization was social media (11.7%, n=50), followed by traditional media sources (6.7%, n=28), and social relations (3.1%, n=13). For example, one user described the way their mother was radicalized after being exposed to misinformation on Facebook and Bichute, an alternative social media platform:

> *"My mom has been watching Bitchute and Facebook videos and fell down the QAnon rabbit hole for the last year, first it was about microchips in the vaccines, a cabal of baby-blood drinking politicians, the great reset. She swallowed it all and regurgitates it with such smugness."*

---

[3]Notably, this finding may not reflect actual Q adherent affiliation, and may not be representative of QAnon demographics.

Only 4.8% (n=20) of the coded posts mentioned the Q relation had deradicalized. One user described the process of deradicalizing their Q father by showing him "centrist" sources, intending to demonstrate that his preferred news sources skewed information to be inflammatory. Another user discussed the way their mother deradicalized when "a nurse, someone she really trusts told her that, 'people are dying every day, unconscious on ventilators, and that the vaccine is what is stopping it." When discussing the state of the relationship between users and their Q relation, 9.6% (n=40) reported discontinuing their relationship with their Q relation, while 4.5% (n=19) reported that their Q relation initiated the end of the relationship. Approximately 2.2% (n=9) of posts reported the decision to discontinue contact was made mutually.

**Table 1**. Frequencies of Coding Themes from Top Posts (n=418)

| Code | Type | Count (%) | Excerpt |
|------|------|-----------|---------|
| Post Type | Validation/ Corroboration | 31 (7.4%) | *I never thought I'd be in this position ever, but I honestly don't know what to do. I created a Reddit account to seek some support so here goes.* |
| | Advice Seeking | 82 (19.6%) | *Do I try to get custody of my kids from my Q?* |
| | Venting | 257 (61.5%) | *Sorry for venting but I'm and sad right now, I appreciate the existence of this community.* |
| Conspiracy | General | 69 (16.5%) | *My Q tried to convince me the earth is flat.* |
| | COVID Vaccine (only) | 118 (28.2%) | *My father is trying to drill into my head the vaccine is out to kill us.* |
| | Both | 135 (32.3%) | *Q believes government is out to control everyone and the vaccine kills people* |
| Covid Vaccine | Pressures family to not vaccinate/use alternative treatments | 157 (37.6%) | *My Q father said to "try some Ivermectin"* |
| | Pressures others | 11 (2.63%) | *I logged into my Q's Facebook, and it was full of antivax posts* |

| Vaccine Reassurance | Asking for Support | 16 (3.8%) | *My parents told me I'd die from the vaccine. If anyone's gotten it please reassure me* |
|---|---|---|---|
| | Giving Support | 3 (0.7%) | *For people who are hesitant, I got the booster and only mild side effects* |
| Covid Infection | Q is sick/got others sick | 34 (8.1%) | *My Q sister got COVID* |
| | Q passed away | 9 (2.2%) | *My electively unvaccinated Q uncle died of COVID* |
| Radicalization | Social media | 50 (11.7%) | *My Q brother gets all his information from Facebook and far-right YouTube videos* |
| | Traditional News | 28 (6.7%) | *My Q reads Newsmax and Fox* |
| | Social Relation | 13 (3.1%) | *My aunt got sucked into QAnon last year and then introduced Q to my parents* |
| Deradicalization | Got Vaccine | 12 (2.9%) | *Whelp, Mom finally got vaccinated! It's been a long campaign of debunking, empathizing, persuading, crying, email propagandizing but by the end she was not actually afraid to get vaccinated* |
| | No Longer Q | 8 (1.9%) | *My Q is thankfully a voracious reader. I noticed he was questioning things and would ask me for evidence. He's now moving away from believing almost everything election/covid related* |
| Disconnect | Q broke connection | 19 (4.5%) | *She finally unfriended me on social media, which I am not at all surprised by. If anything, I am impressed that it actually took this long.* |
| | Poster broke connection | 40 (9.6%) | *So, told my Q our relationship of 20 years was over this week.* |
| | Mutual | 9 (2.2%) | *QAnon has claimed my mother, and ruined my relationship with my dad too. Him and I haven't spoken in over a year behind his QAnon obsession.* |
| Religious | Christian | 31 (7.4%) | *I see Q and radical Christian insanity everywhere i go in my area I'm sick of it* |
| | Atheist | 2 (0.5%) | *My Q husband was once a hippy atheist, but got pulled in by evangelicals , who led him down the rabbit hole of End Times, mixed up with every crazy self-styled prophet and conspiracy theory nutter.* |
| Political | Q is right | 60 (14.4%) | *She has always been...ahem...ultra conservative...* |
| | Q is left | 8 (1.9%) | *We used to joke about how someday that she'd go so far left she'd come back around to the right. Not funny anymore.* |

As seen in **Table 2**, 40% (n=167) of coded posts identified the Q relation(s) as the user's parent(s), 16.3% (n=68) as their partner, 5.3% (n=22) as their friend(s), and 5% (n=21) as their sibling. 23.2% (n=97) of posters did not state their relationship to the Q member.

**Table 2**. Frequencies of Social Relations Mentioned in Top Posts (n=418)

|  | Frequency | Percent |  | Frequency | Percent |
| --- | --- | --- | --- | --- | --- |
| *No relation mentioned* | 97 | 23.2% | *Partner* | 68 | 16.3% |
| *Aunt/Uncle* | 9 | 2.2% | *Professor* | 1 | 0.2% |
| *Child* | 7 | 1.7% | *Self* | 5 | 1.2% |
| *Coworker* | 6 | 1.4% | *Sibling* | 21 | 5% |
| *Family Acquaintance* | 4 | 0.9% | *Stranger* | 5 | 1.2% |
| *Friend* | 22 | 5.3% | *Therapist* | 2 | 0.5% |
| *Grandparent* | 3 | 0.7% | *Uncle* | 1 | 0.2% |
| *Parent* | 167 | 40% |  |  |  |

**Sentiment Analysis of Posts**

Correlations between LIWC post sentiments and number of upvotes and comments were ranked by effect size and are shown in **Table 3**. See appendix **Table A1** for correlations of all LIWC sentiment categories. Posts with higher proportions of words corresponding to swearing, social referents/processes, male references, past-focus, time, physical, all-or-none language, death, and acquiring (i.e., searching for states or goals that serves ones needs) all showed statistically significant (p<.001) positive correlations with the number of post upvotes. Rankings of correlations between sentiments and number of comments were similarly ordered: physical, acquiring, allure, social referents, social, health, time, future-focused, illness, and past-focused. These correlations were all positive and statistically significant (p<.001). Only two sentiments were negatively correlated with both upvote score and number of comments: use of big words (i.e. 7 letters or longer) (r = -.10, p<.001 for upvote; r = -.16, p<.001 for comment) and Analytic language (r = -.12, p<.001 for upvote; r = -.18, p<.001 for comment).

**Table 3.** Top 10 correlations between LIWC sentiment and post engagement (upvote & comment)

| LIWC Sentiment Category | Exemplar words | Upvote Score | LIWC Sentiment Category | Exemplar words | # of Comments |
|---|---|---|---|---|---|
| *Swear* | Sh*t, F*ck | .14 | *Physical* | Medic, Patients | .17 |
| *Social referents* | He, She | .13 | *Acquire* | Get, Take | .17 |
| *Social* | You, We | .13 | *Allure* | Have, Like | .16 |
| *Male* | His, Man | .12 | *Social referents* | He, She | .15 |
| *Past focused* | Was, Had | .12 | *Social* | You, We | .14 |
| *Time* | When, Now | .11 | *Health* | Physician, Patient | .14 |
| *Physical* | Medic, Patients | .11 | *Time* | When, Now | .14 |
| *All/None* | Never, Always | .10 | *Future-focused* | Will, Going to | .14 |
| *Death* | Dead, Die | .10 | *Illness* | Hospital, Sick | .13 |
| *Acquire* | Get, Take | .09 | *Past-focused* | Was, Had | .13 |

*Note*: All correlations are statistically significant at p<.001. Exemplar words are from LIWC 2022 manual.

## Sentiment Analysis of Commenters

In total, 19,643 unique user accounts were associated with all collected comments posted to the subreddit during the study timeframe. On average, each account commented 3.53 times (SD = 9.18) with a median of 1, as shown in **Table 4**. Discounting the Automoderator (a bot that automatically responds to posts), the max number of comments from a single user during the study timeframe was 536. **Figure 1** shows the distribution of comments by author where the data is skewed to the right, showing a power law distribution where a small proportion of users comment at high frequency. The distribution of upvotes by post ($\mu$=659.64, $\sigma$=979.81) appears to be more normally distributed. **Table 5** shows the 10 most common sentiments on average among top commenters and **Table 6** shows examples of comments high in frequently used sentiments. Among users who commented the most, clout (i.e., leadership language) was expressed the most often (62.83% on average). Further, on average, more than 30% of text from active commenters contained words related to authenticity (i.e., perceived honesty), emotional tone, and analytical thinking.

**Table 4**: Summary Statistics of Number of Comments by Average Account

| Statistics | With Automoderator | Without Automoderator |
|------------|--------------------|-----------------------|
| Mean       | 3.76               | 3.53                  |
| Max        | 4482               | 536                   |
| Median     | 1                  | 1                     |
| Std. Dev   | 33.25              | 9.18                  |

**Figure 1**: Distribution of Comments (by Author) and Upvotes (by Post)

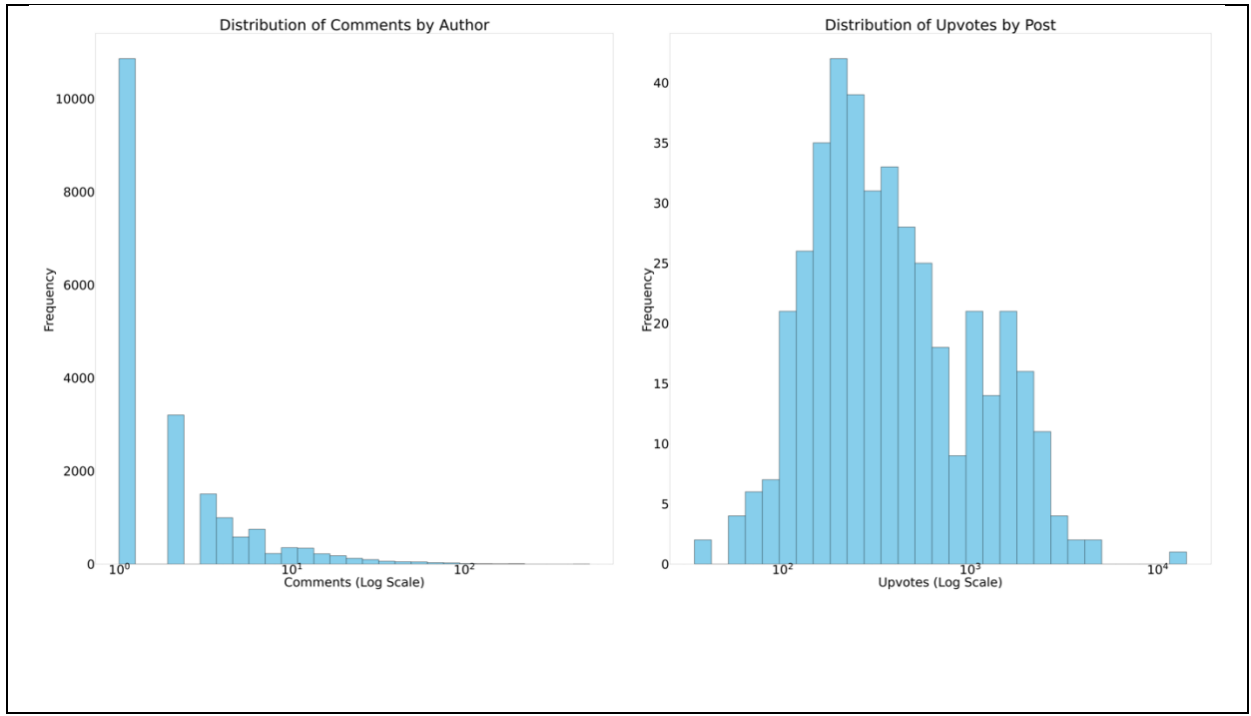

**Table 5**: 10 Most Used Sentiments by Top 10% of Commenter Accounts Ranked by the User's

Average Comment Score

| Most Used Sentiments | | |
|---|---|---|
| *Sentiment* | *Exemplars* | *Average %* |
| Clout | Leadership language, status | 62.83 |
| Authentic | Perceived honesty | 38.35 |
| Tone | Degree of positive (negative) tone | 35.93 |
| Analytic | Metric of logical, formal thinking | 34.6 |
| Social | You, We | 16.5 |
| Cognition | Is, But, Are | 14.18 |
| Cognitive processes | If, Or, Know | 12.75 |
| Social referents | He, She | 11.54 |
| Allure | Have, Like | 7.64 |
| Perception | In, There | 7.60 |

*Note*: Exemplar words are from LIWC 2022 manual. See LIWC manual for further description of sentiment categories.

**Table 6**. Example of Comments with High Percentage of Sentiment

| Comment | Sentiment | LIWC: Description/ Exemplars | Percentage |
|---|---|---|---|
| "If you have a family doctor that's where I would take them. Let them hear it directly from someone who understands." | Clout | Language of leadership, status | 99% |
| "No. When I finally really told my dad how I felt and that I was fine with him until he changes he tripled down and is worse than ever before now." | Authentic | Perceived honesty, genuineness | 99% |
| "Oh, dear. This will not end well. Your days are number. Get an exit plan." | Tone | Degree of positive/negative tone | 99% |
| "he's been having a distorted sense of reality for at least the last 15 years. This is over the pay grade of strangers on the internet. Good luck!" | Analytic | Metric of logical, formal thinking | 99% |
| "Have your sister and mom visit you. Stepdad not invited" | Social | he, she, parent | 70% |
| "That has to be mentally taxing." | Cognitive Progress | but, not, if, or, know | 66.67% |

## Discussion

Findings from this study suggest that users most often turn to the QAnonCasualties subreddit to vent (61.5%, n=257), or to seek advice (19.6%, n=82) or validation (7.4%, n=31). Our results further suggest that Q relations are most often identified as family members: 40% (n=167) of top posts identified their Q relation as a parent, followed by a partner (16.3%, n=68) or a sibling (5%, n=21). Among the most discussed topics, 60.5% (n=253) of the top posts stated that their Q relation(s) were involved with COVID-related conspiracies and 37.6% (n=157) of posts mention pressure to use so-called 'alternative' COVID treatments such as Ivermectin or Hydroxychloroquine, or to not receive treatment or vaccination at all. This finding is of particular concern in cases where users are young adults or minors. Some users sought reassurance about receiving vaccination (3.8%, n=16) and others discussed the repercussions of not vaccinating, as 10.3% (n=43) of top posts noted the Q relation(s) either contracted COVID, gave someone else COVID, or passed away from COVID.

Approximately 16.3% (n=68) of posts explicitly mention disconnecting from Q relation(s). Some of these posts cited specific reasons for disengaging, including fundamental differences and fears of harm from contracting COVID-19. Others describe feeling unable to hold conversations with Q relation(s) without arguing, feeling tired of trying to reason with Q relation(s), or no longer feeling "compatible." These descriptions are consistent with working theories regarding "believer/nonbeliever" relationships, which suggest these relationships may break down due to fundamental differences in perceptions of reality (Jordan and Whitmer, 2023). Though the QAnonCasualties subreddit does not provide a representative sample of the relationships among QAnon adherents and their communities of concern, understanding how and why these relationships break down remains a vital direction for future research. This subreddit, and other public support forums, may provide observational insight.

Descriptions of Q relations tend to align with trends identified in existing research, including higher likelihood of QAnon adherents being Christian and Conservative (DiMaggio, 2022; MacMillen and Rush, 2022). Top posts also included exceptions, with a few users describing Q relation(s) as formerly Liberal (1.9%, n=8) or Atheist (.5%, n=2). Coded posts also included varied descriptions of Q relations, which both match and diverge from the traditional conspiracy characterizations. Given that any descriptions of Q adherents on the QAnonCasualties subreddit are given by friends and family, and do not represent objective measures of personality, our ability to extrapolate based on these posts is limited. However, as QAnon's membership and engagement grows, understanding the composition of Q adherents and the qualities and characteristics that may make individuals susceptible to QAnon remain integral directions for further research.

Commenting by users resembles a power law distribution, where a vast majority of users rarely comment while a small proportion comment frequently. The distribution of post upvotes resembles a normal distribution, suggesting that upvoting occurs more often or consistently than commenting. These findings suggest the average user on QAnonCasualties may prefer "lurking," or passively supporting content via upvoting, while discourse is driven by a small number of highly active users. These dynamics are not unique to the QAnonCasualties community, as previous work suggests social media discourse is often driven by a handful of highly influential accounts (Benevenuto et al., 2009; Grinberg et al., 2019; Haupt, Jinich-Diamant, et al., 2021; Haupt, Li, et al., 2021; Papakyriakopoulos et al., 2020; Romero et al., 2011).

Findings from the sentiment analysis suggest, intuitively, that the QAnonCasualties community responds to content communicating strong emotions, distress, and social situations. Results show that posts containing a higher percentage of swear words and all-or-none (e.g., all, never, always), death (e.g., dead, die), and acquiring (e.g., get, take) words were positively correlated with upvote scores (p<.001). This suggests that users who visit the subreddit may more often upvote content expressing distress or frustration (via swearing and hyperbole). Posts that included social referents (e.g., you, we) and references to males (e.g., his, man) were also positively correlated with upvoting, suggesting that users may engage more often with posts that describe social situations and other people, particularly men. Further, positive correlations between upvotes and percentage of past-focused (e.g., when, then, was, had) and physical (e.g., medic, food, patient) words indicate a higher likelihood of engagement when a post is recounting a past story or concrete incident.

Commenting, which represents a more visible form of online support compared to upvoting, showed similar trends; however, physical and acquire words showed the strongest correlations with commenting, as well as sentiments like health (e.g., physician, patient) and illness (e.g., hospital, sick). This suggests users may be more likely to comment on posts discussing concrete incidents and medically-related topics, such as COVID-19. It is worth noting that these findings are likely influenced by the prevalence and popularity of COVID-19 discussions among top posts during the studied period. Future-focused words (e.g., will, going to) were positively correlated with commenting as well. Since posts with future-oriented language are more likely to describe events that have not occurred yet or express desires for hypothetical outcomes (e.g., "I hope one day"), users may be more likely to interject or give advice by commenting.

Among all examined sentiment categories, only two sentiments were negatively correlated with upvotes and comments: number of big words (7 letters or longer) and analytical words. This finding is consistent with existing research, which suggests properties related to linguistic complexity (such as higher word count and use of big words) decrease the likelihood of post engagement on social media (Davis et al., 2019; Deng et al., 2021; Gligorić et al., 2019; Jones et al., 2004; Temnikova et al., 2015). In sum, users on this subreddit may prefer engaging with posts that express distress, describe concrete incidents and social situations, and discuss events and concerns related to COVID-19, and not posts that use more complex language and analytical themes.

**Recommendations for Cultivating Online Support**

As the popularity of QAnon grows and Q conspiracies continue edging into more mainstream political narratives, understanding the needs of the communities of concern around adherents becomes increasingly vital. Existing research suggests that maintaining relationships with radicalized individuals, or with loved ones who have become fixated on radical beliefs, is associated with higher levels of self-reported anxiety and PTSD symptoms (Moskalenko et al.,

2023). In circumstances where individuals may feel isolated, online support groups have the potential to provide a space to share experiences, seek advice, and find community.

The existence of the QAnonCasualties subreddit suggests the need for an accessible anonymous forum to support the communities of concern around Q adherents. In other words, the subreddit clearly fulfills a necessary function.  In so doing, it provides a window into the needs of these communities.

As discussed, the QAnonCasualties subreddit likely benefits from being both public (open to all users, on a free and accessible platform, etc.) and anonymous (allowing icons and usernames,

not requiring identity verification or personal details, etc.). The public and anonymous nature of

this space also entails drawbacks; with an open platform comes the risk of doxing[4] and trolling,[5]

particularly given that QAnon adherents, and users on its hosting sites 4chan and 8chan, are infamous for online harassment. Other concerns involve the level of effort required to moderate discourse. Reddit moderators are unpaid volunteers, and some have reported struggling with burnout from not having enough time, compensation, or platform support (Dosono and Semaan,

2019; Li et al., 2022; Matias, 2019; Schöpke-Gonzalez et al., 2022). Given the clear utility of online support forums, this study recommends further research into possible avenues of safeguarding and supporting these sites, with the goal of offering greater security while preserving anonymity and accessibility.

[4]The intentional public release of personal information about an individual by a third party, often with the intent to humiliate, threaten, intimidate, or punish the identified individual (Douglas, 2016).
[5]A specific type of malicious online behavior intended to disrupt interactions, aggravate users, or lure them into fruitless arguments (Coles and West, 2016).

Online support groups may also highlight areas of need. During the studied period, users utilized the QAnonCasualties subreddit as a space to vent, seek advice, and seek validation. During this timeframe (7/2021 to 7/2022), principal concerns included the breakdown of relationships with Q relations, COVID-19, and the COVID vaccine. This analysis is retrospective, but current analysis of trends in QAnonCasualties or other public support forums may provide insight into areas of distress or concern for the communities around Q adherents. In other words, our findings suggest public support forums represent important data points for psychological and mental health professionals, offering insight into areas of unmet need.

Finally, though QAnonCasualties represents an independent online community, users may benefit from clearer access to mental health resources. While national and regional resources vary, many health and mental health-related subreddits include "pinned" (permanently available) lists of resources and relevant organizations. Recent research on effective online support groups recommends giving users access to science-backed information to increase access to conspiracy debunking resources (Engel et al., 2023). In instances where users show interest, subject matter experts and mental health professionals may also coordinate with moderators to run IAMAs (I am A __, Ask Me Anything), creating structured opportunities for users to engage with mental health professionals without encroaching on the support space. Overall, platform developers and organizations interested in creating online spaces for healthy support groups should consider providing features that help distressed and vulnerable users access resources, accurate information, and offline physical support if needed.

**Limitations**

This study examines engagement behaviors within the context of QAnonCasualties; as a result, the generalizability of these findings to other online support groups is limited. Further, this study only examines a single support group on a single platform (Reddit). Future work is needed to examine dynamics of other types of support groups and across different online platforms. Moreover, sentiment analysis from LIWC-2022 utilizes NLP techniques that may have trouble picking up the fast-evolving nature of far-right slang. Posters on the subreddit may have referenced specific far-right discourse when referring to their Q relation that LIWC may not have picked up on. Despite these limitations, we believe that these results still provide general insight into the types of topics and emotional sentiment that encourage participation in this online support group.

Additionally, posts report experiences from their authors and may not actually reflect the character or radicalization experience of Q members. Work interested in investigating the background, characteristics, or radicalization process of Q members would require representative data from current or former QAnon members.

**Conclusion**

Online radicalization is an ongoing issue, which represents a growing concern as media becomes more decentralized and users have increased access to more specialized echo chambers. Online support communities may be a promising avenue for communities of concern to receive support. More readily available safeguards and accessible mental health resources may further improve the safety and quality of support that users receive in these environments. Additionally, analysis of popular topics and trends in these spaces may provide insight into unmet needs in these communities, suggesting potential areas to dedicate and invest resources.

**Appendix**

**Table A1**. All bivariate correlations between LIWC sentiment categories and post engagement (upvote and comment)

| *LIWC Sentiment Category* | Upvote | Comm | LIWC | Upvote | Comm | LIWC | Upvote | Comm |
|---|---|---|---|---|---|---|---|---|
| *Word Count* | .034 | .053* | *Swear* | .138** | .109** | *Want* | .042 | .04 |
| *Analytic* | -.118** | -.183** | *Social* | .125** | .143** | *Acquire* | .091** | .165** |
| *Clout* | 0.02 | -.031 | *Socbehav* | .065** | .076** | *Lack* | .043 | .50* |
| *Authentic* | .041 | .122** | *Prosocial* | .015 | .02 | *Fulfill* | .031 | .60* |
| *Tone* | -0.039 | -0.038 | *Polite* | .017 | .01 | *Fatigue* | .065** | .82** |
| *WPS* | .025 | .060* | *Conflict* | .029 | .037 | *Reward* | .004 | .002 |
| *Big Words* | -.099** | -.164** | *Moral* | .060* | .049* | *Risk* | .035 | .053* |
| *Diction* | .110** | .172** | *Comm* | .048 | .057* | *Curiosity* | -.03 | .031 |
| *Drives* | .076** | .090** | *Socrefs* | .128** | .147** | *Allure* | .091** | .155** |
| *Affiliation* | .075** | .080** | *Family* | .064* | .079** | *Perception* | .053* | .111** |
| *Achieve* | .052* | .079* | *Female* | .066** | .085** | *Attention* | -.016 | -.024 |
| *Power* | .02 | .024 | *Male* | .122** | .116** | *Motion* | .068** | .109** |
| *Cognition* | .048 | .101** | *Culture* | -.007 | -.017 | *Space* | .044 | .105** |
| Allnone | .100** | .113** | Politic | .012 | .028 | Visual | .004 | .037 |
| Cogproc | .031 | .087** | Ethnicity | -.009 | -.014 | Auditory | -.009 | -.002 |
| Insight | .017 | .068** | Tech | -.017 | -.041 | Feeling | .045 | .052* |
| Cause | .007 | .050* | Lifestyle | .032 | .060* | Time | .109** | .141** |
| Discrep | .050* | .094** | Leisure | .003 | .002 | Focus past | .121** | .128** |
| Tentative | .008 | .041 | Home | .043 | .056* | Focus present | .050* | .116** |
| Certitude | .012 | .027 | Work | .018 | .036 | Focus future | .073** | .138** |
| Differ | .038 | .080** | Money | .016 | .038 | Conversation | .008 | .034 |
| Tone Pos | .027 | .050* | Religion | .021 | .041 | Net speak | -.002 | .014 |
| Tone Neg | .078** | .093** | Physical | .106** | .168** | Assent | .037 | .072** |
| Emotion | .063** | .076** | Health | .080** | .143** | Nonflu | .038 | .065** |
| Emo Pos | .023 | .03 | Illness | .074** | .129** | Substance | -.002 | .002 |
| Emo Neg | .061* | .074** | Wellness | -.016 | -.005 | Sexual | .012 | .007 |
| Emo Anx | .022 | .086** | Mental | .02 | .008 | Memory | -.003 | -.011 |
| Emo Anger | .027 | .033 | Food | .015 | .034 | Affect | .085** | .1006** |
| Emo Sad | .050* | .048 | Death | .098** | .090** | Need | .075** | .122** |
| | | | Friend | -.021 | .008 | Filler | .005 | .037 |

*Note*: *=p<.05, **=p<.001; green cells indicate positive correlation, red cells indicate negative correlation.